\documentclass[twocolumn]{emulateapj}
\usepackage{graphicx}	% Including figure files
\usepackage{amsmath}	% Advanced maths commands
\usepackage{mathtools}
\usepackage{color}

\newcommand{\grb}{GRB\,170817A}
\newcommand{\host}{NGC\,4993}

\begin{document}

\title{Lessons from the short GRB$\,$170817A -- the First Gravitational Wave Detection of a Binary Neutron Star  Merger}

%\AuthorCallLimit=4
\author{Jonathan Granot\altaffilmark{1}, Dafne Guetta\altaffilmark{2}, and Ramandeep Gill\altaffilmark{1,3}}
\altaffiltext{1}{Department of Natural Sciences, The Open University of Israel, 1 University Road, POB 808, Raanana 4353701, Israel}
\altaffiltext{2}{Department of Physics and Optical Engineering, ORT Braude College, Karmiel 21982, Israel}
\altaffiltext{3}{Physics Department, Ben-Gurion University, P.O.B. 653, Beer-Sheva 84105, Israel}

%\maketitle

\begin{abstract}

The first, long awaited, detection of a gravitational wave (GW) signal from the merger of a binary neutron-star 
(NS-NS) system was finally achieved (GW$\,$170817), and was also accompanied by an electromagnetic counterpart -- the short-duration 
\grb. It occurred in the nearby ($D\approx40\;$Mpc) elliptical galaxy \host, and showed optical, IR and UV emission from half a day up to weeks after the event, as well as late time X-ray (at$\;\geq8.9\;$days) and radio (at$\;\geq16.4\;$days) emission. There was a delay of 
$\Delta{}t\approx1.74\;$s between the GW merger 
chirp signal and the prompt-GRB emission 
onset, and an upper limit of $\theta_{\rm{}obs}<28^\circ$ was set on the viewing angle w.r.t the jet's symmetry axis from the GW signal.
In this letter we examine 
some of the implications of these groundbreaking observations. The delay $\Delta{}t$ sets an upper limit on the 
prompt-GRB emission radius, $R_\gamma\lesssim2c\Delta t/(\theta_{\rm{}obs}-\theta_0)^2$, for a jet with sharp edges at an 
angle $\theta_0<\theta_{\rm{}obs}$. \grb's relatively low isotropic equivalent $\gamma$-ray energy-output may suggest 
a viewing angle slightly outside the jet's sharp edge, $\theta_{\rm{}obs}-\theta_0\sim(0.05-0.1)(\Gamma/100)^{-1}$, but its peak 
$\nu{}F_\nu$ photon energy and afterglow emission suggest instead that the jet does not have sharp edges and the prompt 
emission was dominated by less energetic material along our line of sight, at $\theta_{\rm{}obs}\gtrsim 2\theta_0$. 
Finally, we consider the type of remnant that is produced by the NS-NS merger and find that a relatively long-lived ($>2\;$s) massive NS is strongly disfavored, while a hyper-massive NS of lifetime $\sim1\;$s appears to be somewhat favored over the direct formation of a black hole.

\end{abstract}

%%%%%%%   INTRODUCTION   %%%%%%%%%%%%%%%%%%%%%%%%%%%%%%%%%%%%%%%%%%%%%%%
\section{Introduction}
\label{sec:intro}

The first discovery of gravitational waves (GWs) from two coalescing black holes (BHs) by the 
Advanced Laser Interferometer Gravitational-wave Observatory (LIGO) ushered in a new era of GW 
astronomy \citep{Abbott+16a}. It was soon followed by three other BH-BH mergers
%that detected the inspiral and merger of two BH binaries, 
that firmly established LIGO's sensitivity 
to robustly detect such sources out to $\sim$Gpc distances. 
%Depending on the masses, 
LIGO can also detect GWs from compact binary mergers involving neutron stars (NSs), NS-NS and NS-BH, at a volume-weighted mean distance of 
$\sim\,70\;$Mpc and $\sim\,110\;$Mpc, respectively, and set an upper limit of 
$12,\!600\;{\rm Gpc}^{-3}\,{\rm yr}^{-1}$ on the  NS-NS merger rate  
 \citep[90\% CL;][]{Abbott+16b}.

An electromagnetic (EM) counterpart to the GW signal from a BH-BH merger is not expected (in most scenarios). However, its detection is of great importance in NS-NS or NS-BH mergers, which have been posited to be the progenitors of short-hard gamma-ray bursts 
\citep[SGRBs; e.g.][]{Eichler+89,NPP92}. A NS-NS merger leads to the formation of a BH, possibly preceded by a short-lived 
hyper-massive NS \citep[e.g.][]{Baumgarte+00}. Accretion onto the BH launches a relativistic jet reaching bulk Lorentz factors $\Gamma\gtrsim100$ and powering a 
SGRB -- a short ($\lesssim2\;$s) intense flash of $\gamma$-rays with a typical ($\nu F_\nu$-peak) photon energy $E_{\rm pk}\sim400$~keV and total isotropic-equivalent energy release 
$E_{\gamma,\rm iso}\simeq10^{49}-10^{51}\;$erg \citep[][]{Nakar07,Berger14}. On the other hand, long-soft GRBs are known to originate from the death of massive stars, via   
their association with star-forming regions and type Ic core-collapse supernovae \citep[e.g.,][]{WB06}.

The unprecedented observation of an SGRB \citep{GCN21520,LIGO:ApJ1} coincident with the detection of GWs from coalescing binary NSs \citep{LIGO:PRL} 
in an elliptical galaxy presents the long-awaited ``smoking gun'' that binary NS mergers give rise to SGRBs. Rapid 
follow-up observations by detectors across the EM spectrum \citep{capstone} both increase the positional accuracy of the source in the 
host galaxy and yield critical information regarding jet geometry, merger ejecta, and r-process elements 
\citep[e.g.][]{RPN13}.  
 
In \S$\,$\ref{sec:t_delay} the delay between the GW and SGRB signals is used to 
%derive important inferences regarding 
constrain the location of the $\gamma$-ray emission region. In \S$\,$\ref{sec:prompt} the prompt $\gamma$-ray emission properties are used to constrain the GRB jet's angular structure and our viewing angle $\theta_{\rm obs}$ from the jet's symmetry axis. 
The constraints on the type of remnant produced by the NS-NS merger are discussed in \S\,\ref{sec:remnant}. 
Finally, the implications of this work are discussed in \S$\,$\ref{sec:dis}.

\section{The Time Delay Between the GW Signal and SGRB: An Upper Limit on the  Emission Radius}
\label{sec:t_delay}

A delay of $\Delta t=1.74\pm0.05\;$s was found between the binary merger GW chirp signal and \grb's $\gamma$-ray emission onset \citep{LIGO:ApJ1}.
Such a delay can 
arise from one or more causes, and may provide important information on the merging system and the merger process \citep[e.g.][]{LIGO:ApJ1,IN17,Murguia-Berthier+17}. Moreover, the GW signal and known distance to the host galaxy set an upper limit on the viewing angle of $\theta_{\rm obs}<28^\circ\approx0.49\;$rad \citep{LIGO:PRL}.

One possible cause for such a delay is the formation of a short lived, $t_{\rm HMNS}\lesssim1\;$s, hyper-massive NS, whose collapse 
forms a BH surrounded by an accretion disk that launches a relativistic jet. In order to produce a GRB, the jet must first bore its way through the dynamical ejecta and/or neutrino driven wind that was launched during $t_{\rm HMNS}$, causing a time delay of $t_{\rm bo}$, which may typically be a good fraction of a second \citep[e.g.,][]{MP17,NP17}.     
Once the jet breaks out of this wind or outflow, it quickly accelerates to ultra-relativistic speeds, where compactness arguments suggest that its Lorentz factor during the prompt $\gamma$-ray emission, at a distance of $R_\gamma$ from the central source, is $\Gamma\gtrsim 100$.
The delay $t_r$ in the $\gamma$-ray emission onset for an on-axis observer that is caused by this acceleration phase and a possible coasting phase until the jet reaches $R_\gamma$, due to the jet's motion along the radial direction at speeds slightly less than $c$, is typically negligible (usually $t_r\lesssim R_\gamma/2c\Gamma^2 = 1.7R_{\gamma,13}/\Gamma_{2.5}^2\;$ms\footnote{We adopt the convention $Q_x = Q/10^x$ (c.g.s. units)}).

\begin{figure}[t]
{\par\centering \resizebox*{0.95\columnwidth}{!}
{\includegraphics{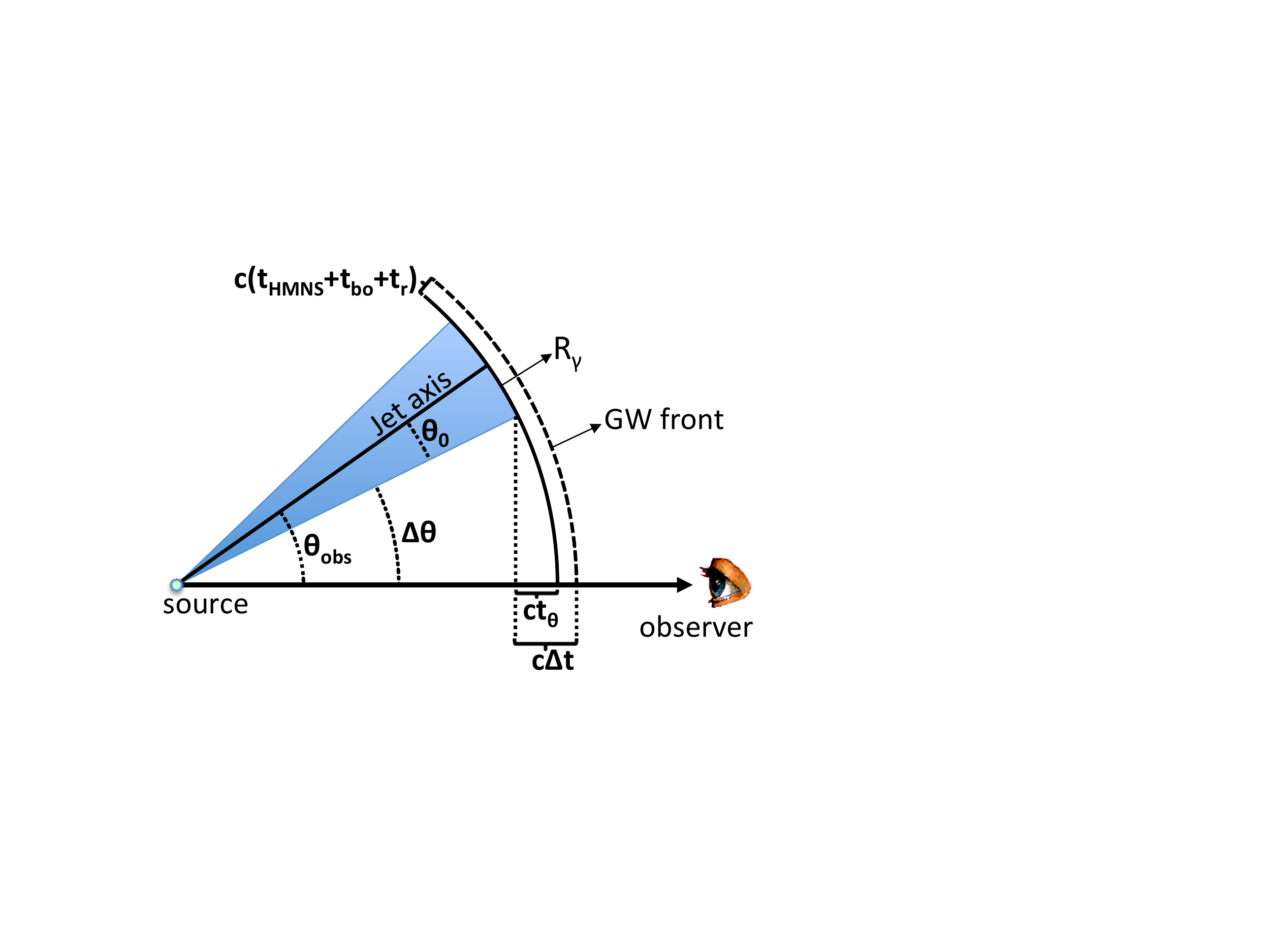}} \par} \caption{\label{fig:time_delay} Illustration of the $\gamma$-ray to GW geometrical time delay $t_\theta$ 
%of the prompt GRB emission 
for a jet viewed from outside of its aperture.
%relative to the GW signal.
}
\end{figure}

\begin{figure}[t]
\centering 
\includegraphics[height=0.65\columnwidth,width=0.99\columnwidth]{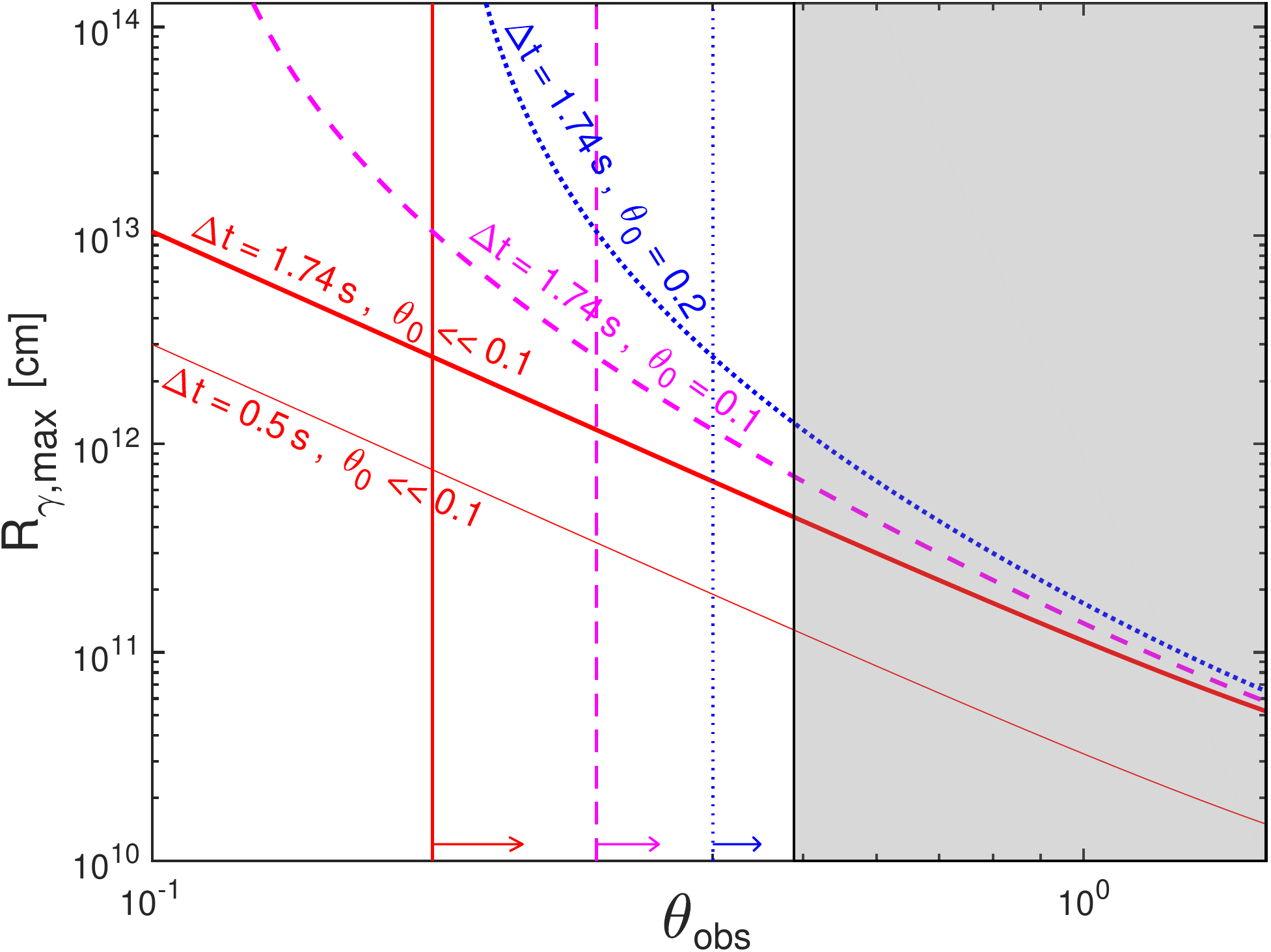}
\caption{ The upper limit $R_{\rm\gamma,max}$ on the $\gamma$-ray emission radius from the geometrical time delay, $t_\theta$, 
%(see Fig.~\ref{fig:time_delay}), 
as a function of the viewing angle, $\theta_{\rm obs}$, for four sets of $(\Delta t,\,\theta_0)$ values. The corresponding vertical 
%dashed black line and two dotted 
lines are tentative lower limits on $\theta_{\rm obs}$ from the fact that afterglow fits suggest $\Delta\theta\gtrsim0.2$.
The gray region is excluded by the GW signal.
}
\label{fig:Rthobs}
\end{figure}

When the outflow in the jet reaches the $\gamma$-ray emission radius, $R_\gamma$, it radiates the prompt GRB. For a jet viewed off-axis from $\theta_{\rm obs}>\theta_0$ this angular offset causes a geometrical delay because of the the additional path length of the radiation from the edge of the jet closest to the observer compared to an on-axis observer (see Fig.~\ref{fig:time_delay}),
\begin{equation}\label{eq:t_theta}
t_\theta = \frac{R_\gamma}{c}\left[1-\cos(\Delta\theta)\right]\approx \frac{R_\gamma}{2c}\Delta\theta^2
= 1.67R_{\gamma,13}\Delta\theta_{-1}^2\;{\rm s}\ ,
\end{equation}
where $\Delta\theta\equiv\theta_{\rm obs}-\theta_0 = 0.1\Delta\theta_{-1}$.
Altogether, the total delay is the sum of all the different causes, $\Delta t\geq t_{\rm HMNS}+t_{\rm bo}+t_r+t_\theta>t_\theta$. Therefore, one can use the fact that $\Delta t>t_\theta\approx R_\gamma\Delta\theta^2/2c$ to set an upper limit on $R_\gamma$,
\begin{equation}\label{eq:R_gamma}
R_\gamma < \frac{c\Delta t}{1-\cos(\Delta\theta)} \approx\frac{2c\Delta t}{\Delta\theta^2} = 6\times10^{12}\left(\frac{\Delta t}{1\;{\rm s}}\right)
\Delta\theta_{-1}^{-2}
%\left(\frac{\theta_{\rm obs}}{0.1}\right)^{-2}
\;{\rm cm}\ .
\end{equation}
%where $\theta_{\rm obs,-1}=\theta_{\rm obs}/0.1$.
This upper limit $R_{\rm\gamma,max}$ on $R_\gamma$ is plotted in Fig.~\ref{fig:Rthobs} for 
$(\Delta t,\,\theta_0) = (0.5\,{\rm s},\,0)$, $(1.74\,{\rm s},\,0)$, $(1.74\,{\rm s},\,0.1)$, $(1.74\,{\rm s},\,0.2)$.
The afterglow lightcurve fits suggest $\Delta\theta\gtrsim0.2$ \citep{Granot17}, as illustrated by the vertical lines in Fig.~\ref{fig:Rthobs}.
Together with the measured $\Delta t$ this implies $R_\gamma\lesssim1.7\times10^{12}(\Delta \theta/0.25)^{-2}\;$cm. Such a limit is very restrictive for models of the GRB prompt emission and outflow acceleration. Note that any estimate or lower limit on the other time delays besides $t_\theta$ that contribute to $\Delta t$ could make this upper limit on $R_\gamma$ even stricter. A long-lived HMNS, $t_{\rm HMNS}\gtrsim0.3-1\;$s for which one might expect $t_{\rm bo}\gtrsim0.5\;$s would imply  a lower $t_\theta\lesssim0.5-1\;$s (see the thin red line in Fig.~\ref{fig:Rthobs} for $t_\theta=0.5\;$s as an illustration of such a case).

\section{Constraining the Viewing Angle from the Prompt GRB Emission}
\label{sec:prompt}
Since the prompt GRB emission was observed, we are either (i) within the jet's initial aperture ($\theta_{\rm obs}<\theta_0$) or beaming cone ($\theta_{\rm obs}<\theta_0+1/\Gamma$), (ii) slightly outside of a sharp edged jet, $\theta_{\rm obs}>\theta_0$ but  $\Gamma\Delta\theta$ is not loo large for the prompt emission to be detectable, or (iii) well outside the core of a jet ($\theta_{\rm obs}\gtrsim2\theta_0$) that (more realistically) does not have very sharp edges and the prompt GRB is produced by relativistic outflow along our line of sight \citep[for further discussion of structured jets and off-axis emission see e.g.][]{Salafia+15,KBG17,Lazzati+17a,Lazzati+17b,Alexander+17,Murguia-Berthier+17,Haggard+17,Jin+17,IN17}.
An alternative that is not discussed here 
in detail is that the prompt GRB and possibly the afterglow arise from the breakout of a mildly relativistic cocoon \citep{Science:Kasliwal,Science:Evans,Nature:Troja,LIGO:ApJ1,Bromberg+17,Gottlieb+17}.

In case (i) 
a bright and usually highly variable prompt $\gamma$-ray emission is expected, with relatively high values of the isotropic equivalent $\gamma$-ray energy output, $E_{\rm\gamma,iso}$, and peak photon energy, $E_{\rm p}$. 
\grb\ had a fluence of $f=(2.8\pm0.2)\!\times\!10^{-7}\;{\rm erg\;cm^{-2}}$ ($10\,$--$\,1000\;$keV) corresponding to $E_{\rm\gamma,iso}=(5.36\pm0.38)\!\times\!10^{46}D_{40\,\rm{Mpc}}^2\;$erg and $E_{\rm p} = 82\pm21\;$keV \citep{GCN21520}. The initial half-second spike had $E_{\rm p} = 185\pm62\;$keV while the softer tail had $E_{\rm p} = 40\pm6\;$keV and a black-body spectrum \citep{capstone}. For SGRBs with known redhsifts 
typically $E_{\rm\gamma,iso}\sim10^{49}$--$10^{51}\;$erg, i.e. 
%$\sim\,10^3\,$--$\,10^4$ times larger than for
$\sim\,3\,$--$\,4$ decades above
 \grb, and an intrinsic $\langle(1+z)E_p\rangle\sim500-600\;$keV, several times larger than in \grb. The low  $E_{\rm\gamma,iso}$ and $E_{\rm p}$ in \grb\  suggest a line of sight is outside of the jet, $\theta_{\rm obs}>\theta_0$, arguing against case (i) above
%(in which $\theta_{\rm obs}\leq\theta_0$). 
($\theta_{\rm obs}>\theta_0$ also correspond to most of the total solid angle for narrow jets and are thus more likely for events associated with a  binary merger GW signal), as do the afterglow observations.

In case (ii) the observed low $E_{\rm\gamma,iso}$ and $E_{\rm p}$ values are caused by a viewing angle outside of the jet's initial aperture, $\theta_{\rm obs}>\theta_0$. 
For a uniform jet with sharp edges
the ratio of off-axis to on-axis $E_{\rm p}$ and $E_{\rm\gamma,iso}$ are \citep{off-axis2002,GRP05,GR12}:   
\begin{eqnarray}
    \frac{E_{\rm p}(\theta_{\rm obs})}{E_{\rm p}(0)} \equiv a &\approx&\left\{
\begin{matrix}1 &\quad \theta_{\rm obs} < \theta_0\ ,\\
\frac{1}{1+(\Gamma\Delta\theta)^2}
\sim (\Gamma\Delta\theta)^{-2} &\quad\theta_{\rm obs}>\theta_0\ ,
\end{matrix}\right.
\\
\frac{E_{\rm\gamma,iso}(\theta_{\rm obs})}{E_{\rm\gamma,iso}(0)} &\approx&\left\{
\begin{matrix}1 &\quad \theta_{\rm obs} < \theta_0\ ,\\
a^2
%\approx[1+\Gamma^2(\theta_{\rm obs}-\theta_0)^2]^{-2}
\sim (\Gamma\Delta\theta)^{-4} &\  1<\frac{\theta_{\rm obs}}{\theta_0}<2\ ,\\
%\frac{(\Gamma\theta_0)^2}{a^{-3}}\approx
\frac{(\Gamma\theta_0)^2}{(\Gamma\Delta\theta)^{6}}
\sim\frac{(\Gamma\theta_0)^2}{(\Gamma\theta_{\rm obs})^{6}} &\  \theta_{\rm obs}>2\theta_0\ ,
\end{matrix}\right.
\end{eqnarray}
where we assume that $\Gamma\theta_0\gg1$, as is inferred for GRBs. 
For $1<\theta_{\rm obs}/\theta_0<2$, $E_{\rm\gamma,iso}(\theta_{\rm obs})/E_{\rm\gamma,iso}(0)\sim (\Gamma\Delta\theta)^{-4}$ already reaches its inferred value of $\sim\,\!10^{-4}\,$--$\,10^{-3}$ for $\Delta\theta\sim (0.05-0.1)(100/\Gamma)\lesssim 0.05-0.1$, where $\Gamma\gtrsim100$ for GRBs, 
i.e. in case (ii) we are observing only slightly outside the jet's outer edge. 
For \grb\ this implies 
$E_{\rm p}(\theta_{\rm obs})/E_{\rm p}(0) = a =[E_{\rm\gamma,iso}(\theta_{\rm obs})/E_{\rm\gamma,iso}(0)]^{1/2}\sim 30-100$ and hence $E_{\rm p}(0)\sim\,\!3\,$--$\,8\;$MeV (or $\sim\,\!5\,$--$\,20\;$MeV for the main half-second initial spike), which is unusually high for an SGRB of typical $E_{\rm\gamma,iso}$. 

Case (iii) allows large off-axis viewing angles $\theta_{\rm obs}\gtrsim 2\theta_0$ for which the afterglow emission from the jet's core peaks and joins the post jet-break on-axis lightcurve at $t_{\rm peak}\propto\theta_{\rm obs}^2$ \citep[e.g.][]{off-axis2002,NPG02}. Moreover, in this case we expect in addition to this off-axis emission from the jet's core also a contribution to the afterglow lightcurve from the material along the line of sight after it produces the prompt GRB. The latter may dominate at early times before $t_{\rm peak}(\theta_{\rm obs})$ while the emission from the jet's energetic core ($\theta<\theta_0$) is still strongly beamed away from the observer.
At $t\gtrsim t_{\rm peak}$ the line of sight enters the beaming cone of the jet's core so that its larger energy causes its emission to dominate over that from the less energetic material along the line of sight at $t\gtrsim t_{\rm peak}$. Therefore, in case (iii) a shallower rise to the peak flux at $t_{\rm peak}$ may be expected \citep[e.g.,][]{off-axis2002,EG06,GK03}.

The early afterglow emission from material along our line of sight in case (iii) may be estimated by assuming spherical emission with the local isotropic equivalent kinetic energy 
$E_{\rm k,iso}\sim E_{\rm\gamma,iso}\approx5.4\times10^{46}D_{40\,\rm{MPc}}^2\;$erg. The latter assumption is reasonable at sufficiently early times when one expects that $E_{\rm k,iso}$ has not greatly changed from its initial value. The flux densities in the relevant power-law segments of the spectrum are \citep[][after the local deceleration time]{GS02}:
\begin{eqnarray}\label{Fnu3a}
F_{\nu>\nu_c,\nu_m}&=& 0.60\epsilon_{e,-1}^{p-1} \epsilon_{B,-2}^\frac{p-2}{4}
t_{\rm days}^{(2-3p)/4}\nu_{14.7}^{-p/2}\ \mu{\rm Jy} \ ,\quad 
\\ \label{Fnu2b}
F_{\nu_m<\nu<\nu_c}&=& 7.6\epsilon_{e,-1}^{p-1} \epsilon_{B,-2}^\frac{p+1}{4}
n_0^{1/2}
t_{\rm days}^\frac{3-3p}{4}\nu_{14.7}^\frac{1-p}{2}\
{\rm nJy}\ ,\quad\ \ 
\\ \label{Fnu3c}
F_{\nu_a<\nu<\nu_m<\nu_c}&=& 156\epsilon_{e,-1}^{-\frac{2}{3}}\epsilon_{B,-2}^{1/3}
n_0^\frac{1}{2}
t_{\rm days}^{1/2}\nu_{9.93}^{1/3}\
\mu{\rm Jy}\ ,\quad
\end{eqnarray}
with the numerical coefficient evaluated for $p=2.5$.
These fiducial values correspond to $\nu F_\nu\approx1.6\times10^{-15}\;{\rm erg\;cm^{-2}\;s^{-1}}$ at $h\nu=1\;$keV for $\nu>\nu_c,\,\nu_m$ after one day, which is consistent with the {\it Chandra} upper limit \citep{Margutti17} of $F_X<1.4\times10^{-15}\;{\rm erg\;cm^{-2}\;s^{-1}}$ ($0.3-10\;$keV) at $2.3\;$days. The corresponding optical magnitude for $\nu_m<\nu<\nu_c$ after one day is $R\sim29$, which would be extremely hard to detect. The radio upper limit of $F_{10\,{\rm GHz}}<15.4\,\mu$Jy at $1.39\;$days \citep{Science:Hallinan} is quite constraining here (a factor of $\approx13$ below the flux from Eq.~\ref{Fnu3c} for our fiducial values), and favors lower values of $n_0$ and/or $\epsilon_B$. Therefore, even if such material along the line of sight produces the observed GRB prompt emission, its afterglow emission would be very challenging to detect.

\section{The Remnant of the NS-NS Merger}
\label{sec:remnant}

\begin{figure}
\centering
\includegraphics[width=0.99\columnwidth]{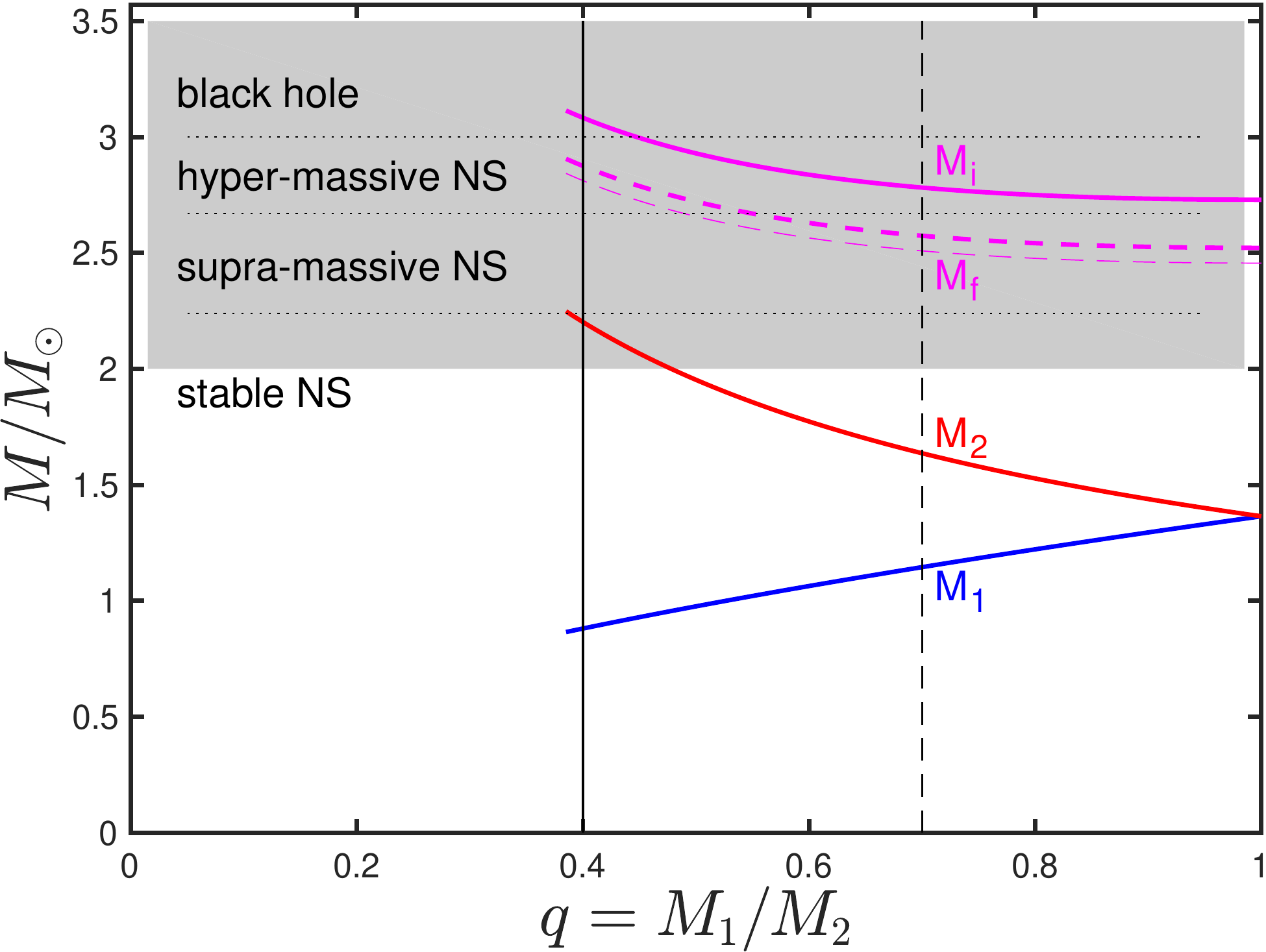} 
\caption{The possible pre-merger masses of the two NSs, $M_1$ ({\it in blue}) and $M_2$ ({\it in red}), as a function of their mass ratio $q\equiv M_1/M_2\leq1$, given the measured chirp mass, $\mathcal{M}\equiv(M_1M_2)^{3/5}(M_1+M_2)^{-1/5} = 1.188_{-0.002}^{+0.004}\,M_\odot$. Also shown are the system's pre-merger total (gravitational) mass, $M_i$, before ({\it solid magenta line}) and after ({\it dashed magenta lines})
accounting for losses due to mass ejection (of $0.01M_\odot$ -- {\it thick line}, or $0.1M_\odot$ -- {\it thin line}), gravity waves and neutrinos during the merger (as described in the text). The vertical thin lines indicate the lower limit on $q$ from the GW signal, for two different priors on the NSs' aligned spin components \citep[$|\chi|\leq0.89$ and $|\chi|\leq0.05$ for the {\it solid} and {\it dashed} lines, respectively; see][]{LIGO:PRL}.
Also shown schematically are the possible outcomes, in order of increasing final mass range: a stable NS, a SMNS, a HMNS, and a BH. The shaded region above $2M_\odot$ indicates the uncertainty in the mass limits dividing the different types of remnants.}
%\vspace{0.2cm}
\label{fig:remnant}
\end{figure}

The type of remnant that was produced during this NS-NS merger is rather uncertain.
The chirp-mass was determined from the GW signal to be $\mathcal{M}\equiv(M_1M_2)^{3/5}(M_1+M_2)^{-1/5} = 1.188_{-0.002}^{+0.004}\,M_\odot$ \citep{LIGO:PRL} where $M_1$ and $M_2$ are the pre-merger (gravitational) masses of the two NSs. Figure~\ref{fig:remnant} shows $M_1$ and $M_2$ as a function of their mass ratio, $q\equiv M_1/M_2\leq1$, along with the initial, pre-merger total mass of the system, $M_{i}=M_1+M_2$.
This measured chirp mass $\mathcal{M}$ implies $M_{i}\geq2.73\,M_\odot$. The final mass, $M_{f}$, of the remnant that was left after the merger can however be smaller \citep[by about $\approx7\%$;][]{Timmes96} due to mass ejection and energy losses to gravitational waves and neutrinos during or shortly after the merger. Therefore, Fig.~\ref{fig:remnant} also shows an estimate of the resulting final mass $M_{f}$ after such a reduction ({\it dashed magenta lines}) by assuming that
$0.01M_\odot$ ({\it thick line}) or $0.1M_\odot$ ({\it thin line})  
of baryonic mass was ejected during the merger and using the relation between the baryonic ($M_b$) and gravitational ($M_g$) masses from \citet{Timmes96}, $M_b = M_g+0.075M_g^2$ in solar masses. 

The merger outcomes can be one of the following, in order of increasing remnant mass: a stable NS, a supra-massive NS 
\citep[SMNS; e.g.][]{Piro+17}, a HMNS, or a BH. A SMNS is supported against collapse to a BH by uniform rigid-body 
rotation, and therefore typically collapses to a BH only on the order of its spindown time $\tau_{\rm sd}$ due to magnetic dipole 
braking. A HMNS is supported against collapse to a BH instead by differential rotation, and has an expected lifetime 
of $t_{\rm HMNS}\lesssim 1\;$s before collapsing to a BH due to the relatively fast damping of differential rotation.

The exact range of final masses that corresponds to each of these outcomes is uncertain and depends on the equation of state (EOS). 
Nonetheless, it is evident that a stable NS remnant in \grb\ would require both approximately equal masses of the merging NSs, as 
well as a very stiff EOS. The GW signal does not strictly rule out the formation of a HMNS \citep{LIGO-PM-GW}, which is the most likely outcome if 
indeed the collapse to a BH was delayed.

A massive NS formed in a binary NS merger would have a near break-up initial spin period of $\lesssim1\;$ms, which corresponds 
to a very large initial rotational energy, $E_{\rm rot}\sim10^{52.5}-10^{53}\;$erg. This energy is channelled primarily into a 
pulsar-type ultra-relativistic MHD wind through magnetic dipole braking, and most of it is lost over $\tau_{\rm sd}$.  For a 
long-lived ($>2\,$s) massive NS remnant, namely a stable NS or a SMNS (and possibly a particularly long-lived HMNS), it is not clear what powers 
the \grb\  since $\tau_{\rm sd}\gtrsim10^2\;$s even for a magnetar-strength magnetic field ($\sim10^{15}\;$G). An even stronger dipole field ($\gtrsim10^{16}\;$G) is needed to give $\tau_{\rm sd}<2\;$s, and then most of $E_{\rm rot}$ would be promptly channeled into the relativistic wind. Moreover, $\tau_{\rm sd}$ does not 
exceed several years even for a typical pulsar-like surface magnetic dipole field strength ($\sim10^{12}\;$G), and therefore by 
the time of the radio to X-ray observations within the first month after the event, at least a few percent of $E_{\rm rot}$, 
i.e. $\gtrsim 10^{51}\;$erg (and possibly most of $E_{\rm rot}$) is extracted. Such an energy in a roughly isotropic relativistic 
wind is expected to produce a very bright afterglow emission as it interacts with the external medium, especially at a nearby 
distance of $\approx40\;$Mpc, which is inconsistent with the multi-wavelength follow-up observations of \grb\ \citep[][]{capstone}. 
Therefore, a stable NS or a SMNS remnant are highly unlikely. 

This leaves either a HMNS or a direct formation of a BH. A HMNS could naturally account for the observed delay $\Delta t$ through $t_{\rm HMNS}$ as well as $t_{\rm bo}$ as the jet would need to bore its way through $\sim1\;$s worth of neutrino-driven wind and dynamical ejecta propagation, whereas for a direct BH formation $t_{\rm HMNS}=0$ and $t_{\rm bo}$ would likely be much shorter as the jet starts very shortly after the disk wind and dynamical ejecta.
A BH would also require a relatively soft EOS. Hence, arguably, 
the most likely option appears to be the formation of an HMNS with a lifetime $t_{\rm HMNS}<\Delta t \approx 1.74\;$s so that its collapse to 
a BH and subsequent accretion onto this BH could launch the jet that powered \grb, and still be consistent with the GRB's delayed 
onset w.r.t the GW chirp signal \citep[also see][for additional arguments favoring the formation of a HMNS]{Margalit-Metzger-2017}.

\section{Discussion}
\label{sec:dis}

We have addressed some implications of \grb\ observations, combining its EM emission, and the associated GW signal of the binary NS merger that triggered it -- the first of its kind. 
In \S$\,$\ref{sec:t_delay} we have used the observed time delay of $\Delta t=1.74\pm0.05\;$s between the GW chirp signal and the GRB onset in order to set an upper limit on the prompt GRB emission radius $R_\gamma\leq R_{\rm\gamma,max}\approx2c\Delta t/\Delta\theta^2$ for a uniform jet with sharp edges viewed from outside of its aperture (see Eq.~(\ref{sec:t_delay}) and Figs.~\ref{fig:time_delay}, \ref{fig:Rthobs}).

Next, in \S$\,$\ref{sec:prompt} we interpreted the relatively low measured values of $E_{\rm\gamma,iso}$ and $E_{\rm p}$ for \grb\ in the context of a narrow GRB jet viewed off-axis, from outside of its initial aperture. For a uniform sharp-edged jet this suggests that our line of sight is only slightly outside of the jet $\Delta\theta\sim (0.05-0.1)(\Gamma/100)^{-1}$, which would in turn imply an unusually high on-axis $E_{\rm p}(0)\sim\,\!3\,$--$\,8\;$MeV (or $\sim\,\!5\,$--$\,20\;$MeV for the main spike), and a relatively high $R_{\rm\gamma,max}$. The implied high $E_{\rm p}(0)$ and the expected afterglow lightcurves both favor an alternative picture (case (iii)), in which our viewing angle is larger ($\theta_{\rm obs}\gtrsim2\theta_0$) and the prompt emission arises from material along our line of sight that is less energetic than the jet's core. This picture implies a higher afterglow flux at very early times, keeps \grb\ well above the Amati relation ($E_{\rm p}\,$--$\,E_{\rm\gamma,iso}$ correlation) like most SGRBs, and induces no angular time delay $t_\theta$. Since the radial time delay $t_r$ is typically negligible for a highly-relativistic outflow, the observed delay $\Delta t\approx1.74\;$s would then likely be dominated by $t_{\rm HMNS}\gtrsim1\;$s and $t_{\rm bo}\sim0.5\;$s. A possible alternative is a mldly relativistic outflow along our line of sight (e.g. from a cocoon that breaks out) for which $\Delta t$ may be dominated by $t_r$.

The latter conclusion is consistent with the arguments raised in \S\,\ref{sec:remnant} against a long-lived massive NS remnant (stable NS or SMNS). While a direct formation of a black hole might still be possible, given the system's expected final mass this would not be the case for many of the leading models for the NS equation of state \citep[e.g.][]{LIGO:ApJ1}. Moreover, it would require another origin for the delay time $\Delta t$, such as the radial time delay $t_r$ for a mildly relativistic outflow.
Nonetheless, the relatively large ejected mass ($\sim0.05M_\odot$) that was inferred from detailed modeling of the kilonova emission 
\citep[e.g][]{Nature:Smartt,Nature:Pian,Nature:Kasen,Science:Drout,Science:Evans,Science:Kasliwal,Science:Kilpatrick} favors relatively small values for the mass ratio, $q\lesssim 0.5-0.6$ for stiff EOSs \citep[e.g.][]{Rosswog14,Sekiguchi16,Dietrich17,Ciolfi17}, which in turn would imply a larger total mass (see Fig.~\ref{fig:remnant}) that would generally more easily lead to a direct BH formation. However the required stiff EOSs for this effect to be important also imply a higher mass threshold for direct BH formation. Finally, if the BH and GRB jet form immediately following the NS-NS merger, then there would be very little neutrino-driven wind in front of the jet's head that would cause a significant fraction of its energy to be channeled into a cocoon, whose breakout might account for such a mildly relativistic outflow along our line of sight. These arguments appear to favor the formation of a short-lived HMNS, with a lifetime of $t_{\rm HMNS}\sim1\;$s or so.

The first detection of a GW signal from the merger of a NS-NS system was observed in coincidence with \grb. We were apparently lucky in the sense that most NS-NS merger GW signals are expected without an associated GRB, since GRB jets are thought to be narrowly collimated, covering only a small fraction, $f_b\sim10^{-2}-10^{-1}$, of the total solid angle. On the other hand, the evidence for narrow jets in short-hard GRBs is much weaker than in long-soft GRBs, so it might be that the beaming factor $f_b$ is larger than expected, which would require us to be somewhat less lucky to have observed the association with \grb. Moreover, it may very well be that our viewing angle $\theta_{\rm obs}$ is not particularly small (since most of the solid angle is at large angles), but the jet does not have sharp edges as is often assumed mainly out of convenience, but instead has wide wings that extend out to large angles from its symmetry axis. In this case the prompt GRB emission in \grb\ was from such material with a low $E_{\rm k,iso}\sim E_{\rm\gamma,iso}$. A determination of $\theta_{\rm obs}$ from the GW signal together with elaborate multi-wavelength afterglow observations could help determine the GRB jet's angular structure, as well as constrain the prompt GRB emission radius $R_\gamma$.

\acknowledgements
JG and RG are supported by the Israeli Science Foundation under Grant No. 719/14. RG is supported by an Open University of Israel Research 
Fellowship.

%\newpage

\end{document}